\documentclass[runningheads]{llncs}
\usepackage[T1]{fontenc}
\usepackage{graphicx}
\usepackage{amsmath,amssymb,amsfonts}
\usepackage{algorithmic}
\usepackage[linesnumbered,ruled]{algorithm2e}

\usepackage{xcolor}
\usepackage{multirow}
\usepackage{booktabs}
\usepackage{bm}
\usepackage{gensymb}
\newcommand*{\eg}{\emph{e.g.}}
\newcommand*{\ie}{\emph{i.e.}}

\usepackage[pagebackref,breaklinks,colorlinks]{hyperref}
\usepackage[capitalize]{cleveref}
\crefname{section}{Sec.}{Secs.}
\Crefname{section}{Section}{Sections}
\crefname{table}{Tab.}{Tabs.}
\Crefname{table}{Table}{Tables}
\crefname{equation}{Eq.}{Eqs.}
\Crefname{equation}{Equation.}{Equations.}

\begin{document}
\title{3D-TransUNet for Brain Metastases Segmentation in the BraTS2023 Challenge}
\titlerunning{3D-TransUNet for BraTS-METS 2023}
%
\author{Siwei Yang\inst{1} \and
Xianhang Li\inst{1}  \and
Jieru Mei\inst{2}\and
Jieneng Chen\inst{2}\and \\ 
Cihang Xie\inst{1} \and
Yuyin Zhou\inst{1}
}

\authorrunning{S. Yang et al.}

\institute{
University of California, Santa Cruz \and
The Johns Hopkins University
}
\maketitle              
\begin{abstract}
Segmenting brain tumors is complex due to their diverse appearances and scales. Brain metastases, the most common type of brain tumor, are a frequent complication of cancer.
Therefore, an effective segmentation model for brain metastases must adeptly capture local intricacies to delineate small tumor regions while also integrating global context to understand broader scan features. The TransUNet model, which combines Transformer self-attention with U-Net's localized information, emerges as a promising solution for this task.
In this report, we address brain metastases segmentation by training the 3D-TransUNet~\cite{chen2023transunet3d} model on the Brain Tumor Segmentation (BraTS-METS) 2023 challenge dataset. Specifically, we explored two architectural configurations: the \textbf{Encoder-only 3D-TransUNet}, employing Transformers solely in the encoder, and the \textbf{Decoder-only 3D-TransUNet}, utilizing Transformers exclusively in the decoder.
For Encoder-only 3D-TransUNet, we note that Masked-Autoencoder pre-training is required for a better initialization of the Transformer Encoder and thus accelerates the training process.

We identify that the Decoder-only 3D-TransUNet model should offer enhanced efficacy in the segmentation of brain metastases, as indicated by our 5-fold cross-validation on the training set\footnote{The code and models are available at~\url{https://github.com/Beckschen/3D-TransUNet}}. However, our use of the Encoder-only 3D-TransUNet model already yield notable results, with an average lesion-wise Dice score of 59.8\% on the test set, securing second place in the BraTS-METS 2023 challenge.

\keywords{Brain Tumor Segmentation  \and Transformer}
\end{abstract}
\section{Introduction}

Tumors, with their subtle intensity variations compared to surrounding tissues, often pose difficulties, as evidenced by inconsistencies in even expert-driven manual annotations\cite{fathi2023automated,wang2021annotation,greenwald2022whole,kazerooni2024brain,adewole2023brain}. Additionally, the wide variance in tumor appearances and dimensions across patients challenges the efficacy of traditional shape and location models\cite{ma2024segment,renard2020variability}.
Brain metastases, which are brain tumors that originate from primary cancers elsewhere in the body, represent the most prevalent malignant tumors in the central nervous system. With an annual incidence of 24 per 100,000 individuals~\cite{habbous2021incidence,boire2020brain,moawad2023brain4}, brain metastases outnumber the occurrence of all primary brain cancers combined.

In terms of segmentation methods, Convolutional Neural Networks (CNNs), especially Fully Convolutional Networks (FCNs)\cite{long2015fully}, have established their prominence. Among various architectures, the u-shaped architecture, popularly known as U-Net~\cite{ronneberger2015u}, stands out for its symmetrical encoder-decoder framework and skip-connections, excelling at preserving image intricacies.
However, these methods often struggle with modeling long-range dependencies due to convolution's inherent locality. To address this, researchers have turned to Transformers, which rely solely on attention mechanisms, showcasing success in capturing global contexts~\cite{vaswani2017attention}. An example is TransUNet~\cite{chen2021TransUNet}, a hybrid CNN-Transformer model, seamlessly blending localized convolution's efficiency with global attention's comprehension. Anchored in the encoder-decoder paradigm, this innovation leverages and elevates both paradigms, promising a new frontier in segmentation precision.

This report aims to validate the performance of 3D-TransUNet~\cite {chen2023transunet3d} on the segmentation of brain metastases in the BraTS 2023 challenge. 3D-TransUNet has two opted self-attention modules: 1) A \emph{Transformer encoder}, which tokenizes image patches extracted from CNN feature maps to capture extensive global contexts using transformer blocks, and 2) A \emph{Transformer decoder}, which innovatively redefines the process of medical image segmentation by treating it as a mask classification task and dynamically refining organ queries through cross-attention with multi-scale CNN decoding features.
Specifically, we experiment with two architectures: \textbf{Encoder-only} (CNN encoder + Transformer encoder + CNN decoder) and \textbf{Decoder-only} (CNN encoder + Transformer decoder + CNN decoder).
Notably, Masked-Autoencoder (MAE) Pre-training can be used to accelerate the training of the Encoder-only model.
To introduce stronger supervision, we employ deep supervision across all levels of our decoder.
Our model yielded average lesion-wise Dice scores of 59.6\% and 59.8\%, respectively, on the validation set and test set of BraTS-METS 2023 datasets.

\section{Method}
\begin{figure*}[t!]
    \centering
    \includegraphics[width=\textwidth]{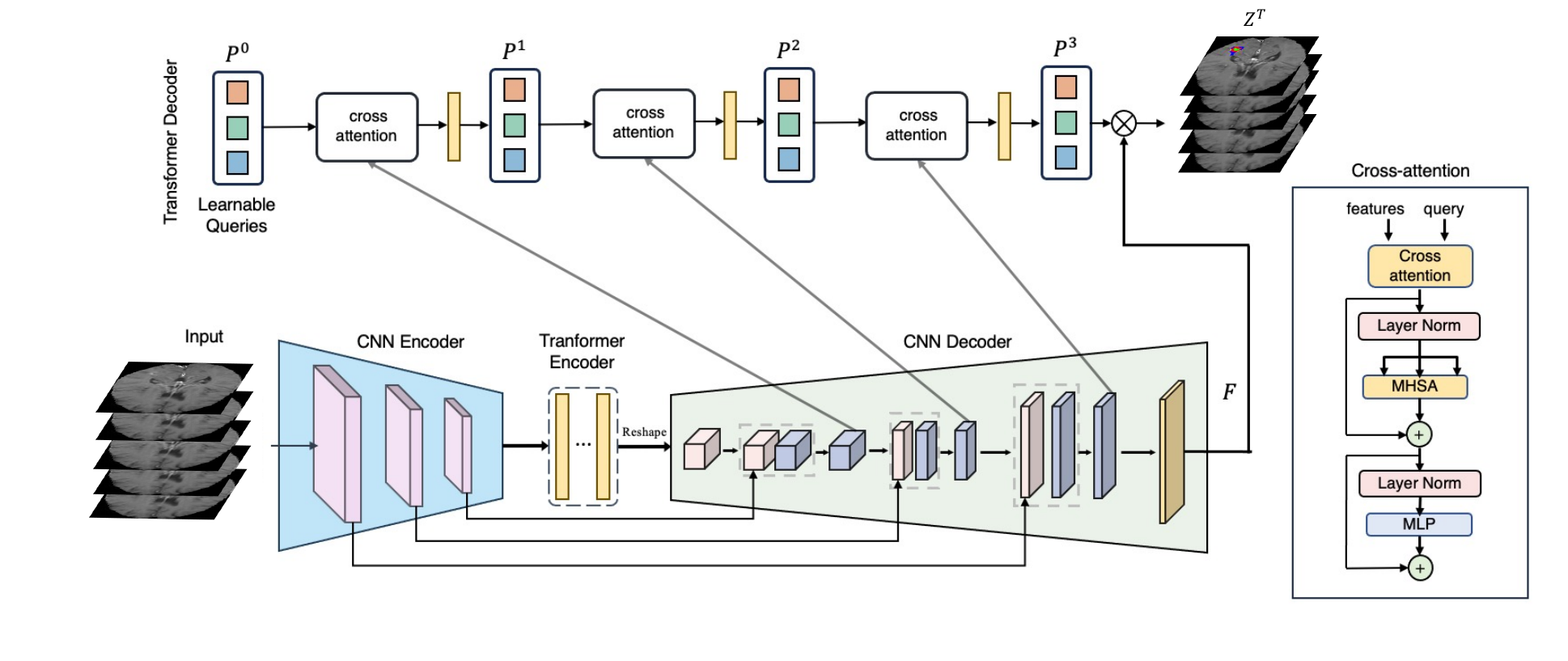}
    \caption{Overview of our adaptation of 3D-TransUNet~\cite{chen2023transunet3d} for BraTS-METS 2023~\cite{adewole2023brain2}. }
    \label{fig:framework}
\end{figure*}

In this work, we adopt 3D-TransUNet~\cite{chen2023transunet3d} to segment brain metastases. 
This model leverages the advantages of integrating transformers within the encoder and decoder of the U-Net architecture, as shown in Figure \ref{fig:framework}. We first studied the encoder to verify if transformer blocks can extract representative features. We also explore combining the U-Net pixel decoder with a Transformer decoder for prediction. The U-Net pixel decoder upsamples the low-resolution features generated by the image encoder. Simultaneously, the Transformer decoder enhances these features through a cross-attention mechanism, effectively refining the final prediction.

\subsection{Transformer as Encoder}
\subsubsection{Image Sequentialization}
An input feature map $\bm{\mathrm{x}}$ are first tokenized and reshaped into a sequence of 3D patches, noted as \{$\bm{\mathrm{x}}^p_i \in \mathbb{R}^{P^3 \cdot C}|i=1,..,N\}$.
The size of each 3D patch is $P \times P \times P$, and the total patch number is $N=\frac{DHW}{P^3}$.

\subsubsection{Patch Embedding}
3D patches $\bm{\mathrm{x}}^p$ are linearly projected into a $d_{enc}$-dimensional embedding space. Learnable positional embeddings are added to retain spatial information. The final embeddings are formulated as follows:

\begin{align}
    \bm{\mathrm{z}}_0 &= [\bm{\mathrm{x}}^p_1 \bm{\mathrm{E}}; \, \bm{\mathrm{x}}^p_2 \bm{\mathrm{E}}; \cdots; \, \bm{\mathrm{x}}^{p}_N \bm{\mathrm{E}} ] + \bm{\mathrm{E}}^{pos}, \label{eq:embedding} 
\end{align}

\noindent where $\bm{\mathrm{E}} \in \mathbb{R}^{(P^2  C) \times d}$ and $\bm{\mathrm{E}}^{pos}  \in \mathbb{R}^{N \times d}$ denotes the linear projection and position embedding accordingly.

The Transformer encoder consists of $L_\textup{enc}$ layers of Multi-head Self-Attention (MSA)~\Cref{eq:msa} and Multi-Layer Perceptron (MLP) blocks~\Cref{eq:mlp}. Therefore, the final output $\bm{\mathrm{z}}_\ell$ of the $\ell$-th layer is

\begin{align}
    \bm{\mathrm{z}}^\prime_\ell &= \textmd{MSA}(\textmd{LN}(\bm{\mathrm{z}}_{\ell-1})) + \bm{\mathrm{z}}_{\ell-1}, &&  \label{eq:msa} \\
    \bm{\mathrm{z}}_\ell &= \textmd{MLP}(\textmd{LN}(\bm{\mathrm{z}}^\prime_{\ell})) + \bm{\mathrm{z}}^\prime_{\ell},   \label{eq:mlp} 
\end{align}

where $\textmd{LN}(\cdot)$ is layer normalization operator and $\bm{\mathrm{z}}_\ell$ is the encoded image representation.

\subsection{Transformer as Decoder}
\label{sec:transformer_decoder}

The segmentation task can be reformulated into a binary mask classification problem inspired by the set prediction mechanism proposed in DETR~\cite{carion2020end}.
As shown in Figure.~\ref{fig:framework}, we train the CNN decoder and the Transformer decoder simultaneously, allowing for the refinement of organ queries and feature maps. 
Specifically,
in the \(t\)-th layer of the Transformer decoder, the refined organ queries are denoted by \(\mathbf{P}^t \in \mathbb{R}^{N\times d_{dec}}\). Alongside, an intermediate feature from the U-Net is transformed into a \(d_{dec}\)-dimensional feature, represented by \(\mathcal{F}\).
The number of upsampling blocks in the CNN decoder aligns with the Transformer decoder layers, so multi-scale CNN features are effectively projected into the feature space \(\mathcal{F} \in \mathbb{R}^{(D_t H_t W_t) \times d_{dec}}\), where \(D_t\), \(H_t\), and \(W_t\) define the spatial dimensions of the feature map at the \(t\)-th upsampling block.
Transitioning from the \(t\)-th to the \(t+1\)-th layer, the organ queries \(\mathbf{P}^t\) are updated through the cross-attention mechanism as described by the following formula:
\begin{align}
\mathbf{P}^{t+1} = \mathbf{P}^{t} + \text{Softmax}\left((\mathbf{P}^{t}\mathbf{w}_{q})(\mathcal{F}^t\mathbf{w}_{k})^\top\right) \mathcal{F}^t\mathbf{w}_{v},
\end{align}
where \(\mathbf{w}_{q} \in \mathbb{R}^{d_{dec} \times d_q}\), \(\mathbf{w}_{k} \in \mathbb{R}^{d_{dec} \times d_k}\), and \(\mathbf{w}_{v} \in \mathbb{R}^{d_{dec} \times d_v}\) are the weight matrices that linearly project the \(t\)-th query features, keys, and values for the subsequent layer. This process is repeated, with a residual connection updating \(\mathbf{P}\) after each layer, in line with the previous method (\cite{cheng2022masked}).
The final prediction, $\mathbf{Z}^T$, is derived through Equation \ref{eq:coarse_prediction}, which details the process of converting $\mathbf{P}^T$ into the binarized segmentation map. It involves a dot product with U-Net's last block feature, $\mathbf{F}$, resulting in $\mathbf{Z}^T$.
\begin{align}
\label{eq:coarse_prediction}
& \mathbf{Z}^{T} = g(\mathbf{P} \times \mathbf{F}^\top),
\end{align}
where $g(\cdot)$ is sigmoid activation followed by a hard thresholding operation with a threshold set at 0.5, such to decode region-wise binary brain tumor masks.

Note that unlike~\cite{chen2023transunet3d}, we do not use masked attention here due to the observed training instability.

\subsection{3D-TransUNet Variants}
\label{sec:variants}
Two 3D-TransUNet variants, \ie, Encoder-only and Decoder-only, are involved in the experiment. Encoder-only 3D-TransUNet is used as the main architecture for all of our submissions since Decoder-only 3D-TransUNet requires longer training compared to the Encoder-only 3D-TransUNet. This is mainly due to its use of high-resolution features in the Transformer decoder. Additionally, the Hungarian matching process used to match binary masks to ground truth in the Decoder-only model is slower compared to directly computing cross-entropy and Dice loss.

\subsubsection{Encoder-Only}
The Transformer encoder along with the CNN encoder compose a CNN-Transformer hybrid encoder in this variant. Feature maps are first extracted by CNN then patchified and tokenized before being fed to the Transformer encoder. A standard U-Net is used as the decoder without the Transformer decoder.

\subsubsection{Decoder-Only}
This variant uses a conventional CNN encoder only for the encoding phase while both CNN and Transformer are used as the decoder. Before being processed by the Transformer decoder, they are augmented with learnable positional embeddings following Eq.~\eqref{eq:embedding}.

\subsection{Training Details}
\label{sec:training}

\vspace{1ex}\noindent\textbf{Masked-Autoencoder (MAE) Pre-training.}
For our \textbf{Encoder-only 3D-TransUNet}, we first pre-train the transformer encoder in an MAE style \cite{he2022masked}.
Specifically,
We initially tokenized 3D input using a 2D patch embedding layer along the z-axis, then flattened it into a 1D sequence.
We randomly mask out 75\% of the tokens and then utilize just one lightweight decoder block to predict the masked tokens. Following \cite{he2022masked}, we calculate the reconstruction loss $L_{\text{recon}} = \frac{1}{N} \sum_{i=1}^{N} (x_i - \hat{x}_{i,\text{masked}})^2$, where \(L_{\text{recon}}\) is the mean square error, \(x_i\) are the original pixel values, \(\hat{x}_{i,\text{masked}}\) denotes the predicted pixel values for the masked tokens, and \(N\) represents the total count of masked tokens.
Following \cite{he2022masked}, we adopt pixel normalization on each patch. 
Specifically, we compute the mean and standard deviation of all pixels in a patch and use them to normalize this patch.
After pre-training, we discard the patch embedding layer and decoder block and solely initialize all transformer encoder blocks with the pre-trained weights.
We observe that utilizing a pre-trained encoder significantly speedup model convergence. 
For instance, with MAE pre-training, the model attains an average  Dice score across five folds of 58.2\% with only 300 epochs of training, whereas a model trained from scratch needs 600 epochs to achieve comparable performance.

\vspace{1ex}\noindent\textbf{Training Loss.}
Unless indicated otherwise, we mainly follow the training details of 3D-TransUNet~\cite{chen2023transunet3d}. In addressing the challenge posed by setting the number of coarse candidates \(N\) considerably greater than the class count \(K\), it becomes inevitable that predictions for each class will exhibit false positives. To mitigate this, we employ a post-processing step to refine the coarse candidates, drawing on a matching process between predicted and ground truth segmentation masks. Taking cues from prior work~\cite{carion2020end,wang2021max}, we utilize the Hungarian matching approach to establish the correspondence between predictions and ground-truth segments. The resulting matching loss is formulated as follows:
\begin{align}
\label{eq:matching_criterion}
\mathcal{L} = \lambda_{0}(\mathcal{L}_{ce} + \mathcal{L}_{dice)} + \lambda_{1}\mathcal{L}_{cls},
\end{align}
Here, the pixel-wise losses \(\mathcal{L}_{ce}\) and \(\mathcal{L}_{dice}\) denote the binary cross-entropy and dice loss~\cite{milletari2016v} respectively, while \(\mathcal{L}_{cls}\) represents the classification loss computed using the cross-entropy for each candidate region. The hyper-parameters \(\lambda_{0}\) and \(\lambda_{1}\) serve to strike a balance between per-pixel segmentation and mask classification loss.

\vspace{1ex}\noindent\textbf{Deep Supervision.}
To introduce stronger supervision, every intermediate level of the 3D-TransUNet decoder produces a prediction map on which the loss function is applied during training.

\section{Experiments and results}

\subsection{Experimental Setting}

\vspace{1ex}\noindent\textbf{Implementation Details.}
All experiments are conducted with a single NVIDIA A5000. Batch size and base learning rate are set as 2 and 2e-3 accordingly. The learning rate follows polynomial decay with a power factor of 0.9.  Augmentation including random rotation, scaling, flipping, white Gaussian noise,
Gaussian blurring, color jittering, low-resolution simulation, Gamma transformation.
Our main experiments in~\Cref{tab:val_test} are conducted using Encoder-only 3D-TransUNet.
The architecture combines a 3D nn-UNet with a pre-trained 12-layer Vision Transformer (ViT) as the Transformer encoder, utilizing Masked Autoencoder (MAE) weights. The latent dimension $d$ is set at 768.
For decoder-only,
the number of layers is 3. And $d_{dec}$ is set to 192.
For training loss \ref{eq:matching_criterion}, $\lambda_{0}$ and $\lambda_{1}$ are set as 0.7 and 0.3.
During the testing phase, we train our model on the entire training set for 600 epochs and submitt the predictions on the validation set. 
During the testing phase, we applied 10-fold cross-validation, where we trained an individual model for every fold for 1,000 epochs.

\vspace{1ex}\noindent\textbf{MAE pretraining settings.}
The input has a shape of $128\times128\times128$ after random cropping.
We train all data on 8 GPUs distributedly for our MAE training, with a batch size of 2 on each GPU.  We train the model in 4800 epochs, including a warm-up period of 40 epochs. The base learning rate is set to 1.5e-4, accompanied by a weight decay 0.05. AdamW optimizer is used by default.

\vspace{1ex}\noindent\textbf{Datasets.}
We report results on BraTS-MET 2023 \cite{moawad2023brain4} which is pivotal for crafting sophisticated algorithms to detect and segment brain metastases, aiming for easy clinical integration. This dataset encompasses a collection of untreated brain metastases mpMRI scans, sourced from multiple institutions and conducted under regular clinical protocols. It should be noted that we didn't use other officially allowed datasets, \eg, NYUMets~\cite{oermann2023longitudinal}, UCSF-BMSR~\cite{rudie2023university}, BrainMetsShare~\cite{grovik2020deep} as these datasets don't share the same mpMRI modalities and annotation format as the BraTS-MET 2023.

It should be noted that we only use BraTS-METS 2023 for training as these datasets share the mpMRI modalities and annotation format.

\vspace{1ex}\noindent\textbf{Evaluation Metrics.}
In assessing the accuracy of a medical image segmentation model, various metrics offer insights into different aspects of performance:

\begin{enumerate}
    \item \textbf{Lesion-wise Dice Score}: Dice score represents the similarity between two binary segmentations. It is given by:
    \begin{equation}
        \text{Dice}(\mathbf{A}, \mathbf{B}) = \frac{2|\mathbf{A} \cap \mathbf{B}|}{|\mathbf{A}| + |\mathbf{B}|}
    \end{equation}
    where \(\mathbf{A}\) and \(\mathbf{B}\) denote the binary segmentations.

    In this challenge, dice scores localized to individual lesions are adopted for lesion-specific evaluation. The Lesion-wise Dice for a designated lesion is:
    \begin{equation}
        \text{Dice}_{\text{lesion-wise}}(\mathbf{A}_i, \mathbf{B}_i) = \frac{2|\mathbf{A}_i \cap \mathbf{B}_i|}{|\mathbf{A}_i| + |\mathbf{B}_i|}
    \end{equation}
    This formula assesses the overlap between the \(i^{th}\) ground-truth lesion \(\mathbf{B}_i\) and all the lesions that overlap with it \(\mathbf{A}_i\).

    \item \textbf{Hausdorff Distance (95\%)}: A metric quantifying the maximum of minimum distances between two binary images at the $95$-th percentile to mitigate outlier effects. Mathematically:
    \begin{equation}
        H(A, B) = \max\left\{\ \sup_{a \in A} \inf_{b \in B} d(a, b), \sup_{b \in B} \inf_{a \in A} d(a, b)\ \right\}
    \end{equation}

\end{enumerate}

\vspace{1ex}\noindent\textbf{Model Ensemble and Test-Time Augmentation}
During testing phase, five models are randomly chosen from ten models trained on each fold for ensemble. Predictions from these five models were averaged to ensemble predictions. To further boost the model's performance with test-time augmentation, predictions from augmented views with flipping and rotation (90\degree, 180\degree, 270\degree) are averaged to produce the final predictions.

\begin{table}[!t]
    \centering
    \begin{tabular}{@{}lcccccccc@{}}
        \toprule
        \multirow{2}{*}{Method} & \multicolumn{4}{c}{Lesion-wise Dice ($\uparrow$)} & \multicolumn{4}{c}{HD95 ($\downarrow$)} \\
        \cmidrule(r){2-9}
        & ET & TC & WT & Avg. & ET & TC & WT & Avg. \\
        \midrule
        Encoder-only 3D-TransUNet~\cite{chen2023transunet3d} & 54.79\%  & 58.96\%  & 56.05\%  & 56.60\%  & 108.9 & 107.6 & 109.5 & 108.7 \\
        Decoder-only 3D-TransUNet~\cite{chen2023transunet3d} & 56.80\%  & 61.12\%  & 60.09\%  & 59.34\%  & 99.4 & 95.9 & 93.97 & 96.4 \\ 
        \bottomrule
    \end{tabular}
    \caption{Ablation Performance of 3D-TransUNet on the training set of BraTS-METS 2023~\cite{moawad2023brain4} under 5-fold cross-validation.}
    \label{tab:met}
\end{table}

\subsection{Encoder-only v.s. Decoder-only}


\begin{table}[!t]
    \centering
    \begin{tabular}{@{}ccccccccc@{}}
        \toprule
        \multirow{2}{*}{Dataset Split} & \multicolumn{4}{c}{Lesion-wise Dice ($\uparrow$)} & \multicolumn{4}{c}{HD95 ($\downarrow$)} \\
        \cmidrule(r){2-9}
        & ET & TC & WT & Avg. & ET & TC & WT & Avg. \\
        \midrule
        Validation & 59.2\%  & 63.4\%  & 56.5\% & 59.6\% & 94.8 & 94.8 & 110.9 & 100.1 \\
        Test & 57.4\%  & 62.0\%  & 59.9\% & 59.8\%  & 103.0 & 99.8 & 99.6 & 100.8 \\ 
        \bottomrule
    \end{tabular}
    \caption{Performance of Encoder-only 3D-TransUNet on the validation and test set of BraTS-METS 2023~\cite{moawad2023brain4}. }
    \label{tab:val_test}
\end{table}

    In order to compare the effectiveness between the Encoder-only model and the Decoder-only model, we apply 5-fold cross-validation on the entire 238 training cases and report the average Dice and HD95 for all testing cases. The 5 models from the 5 folds of BraTS-METS 2023 training set are trained for 200 epochs.
    In \Cref{tab:met}, we report the comparison between Encoder-only 3D-TransUNet and Decoder-only 3D-TransUNet on BraTS-METS 2023's training set with 5-fold cross-validation are presented in \Cref{tab:met}. 
    Compared to our internal validation results with the baseline nnUNet, which yielded average Dice scores of 54.90\%, 58.67\%, and 55.75\% for segmenting ET, TC, and WT, respectively, resulting in an overall average Dice score of 56.44\%, it becomes evident that while the Encoder-only model offers only marginal improvement in segmentation, the Decoder-only model demonstrates a substantial increase in Dice score by 2.9\%. It is important to note that MAE pretraining was not applied to the Encoder-only model in this evaluation. However, with MAE pretraining, the advantages of the Encoder-only model are expected to be more pronounced, albeit still inferior to the Decoder-only model.

\subsection{Main Results}
\begin{figure*}[t!]
    \centering
    \includegraphics[width=0.9\textwidth]{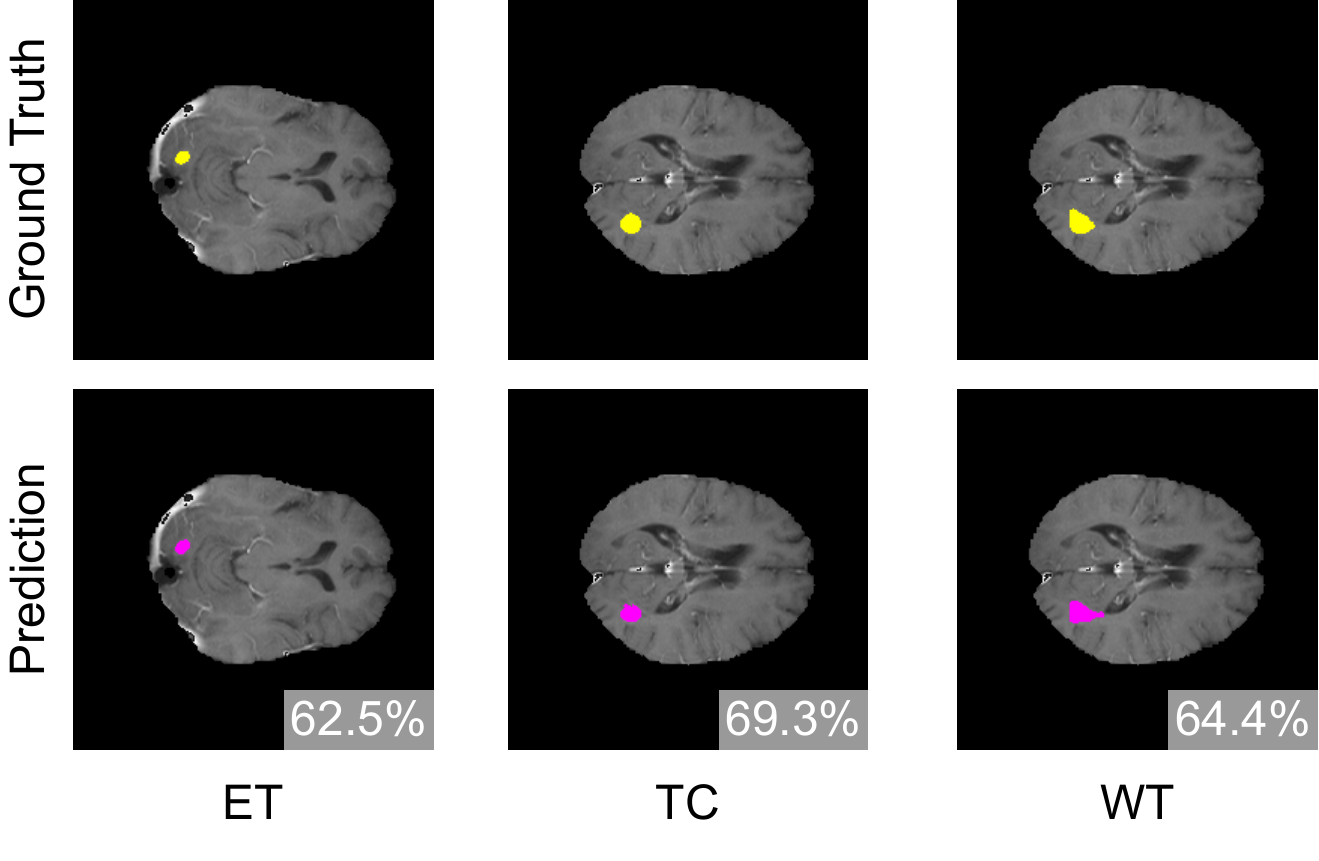}
    \caption{Visual comparison between ground-truths and predictions from Encoder-only 3D-TransUNet on BraTS-METS 2023. The lesion-wise Dice score of each category on this patient is also listed (best viewed in color).}
    \label{fig:met_results}
\end{figure*}

    Since the Decoder-only 3D-TransUNet requires longer training, due to the time and computation limit, we were only able to submit results from the Encoder-only 3D-TransUNet during the validation phase and testing phase, where the official evaluation results are presented in \Cref{tab:val_test}. Specifically, we achieve average lesion-wise Dice scores of 59.6\% and 59.8\% on the validation and test set, securing the second place in the BraTS 2023 challenge.
   We also display a qualitative example to  further demonstrate our method's effectiveness, as shown in \cref{fig:met_results}.


\section{Conclusion}
Brain tumors, especially brain metastases, present challenges in segmentation due to their diverse appearances and sizes. The TransUNet model, combining Transformer self-attention and U-Net's features, shows promise for this task. We trained the 3D-TransUNet model on the BraTS-METS 2023 dataset for brain metastases segmentation, exploring Encoder-only and Decoder-only configurations. Pre-training the Encoder-only model with Masked-Autoencoder improves initialization, facilitating faster training. Although the Decoder-only model is expected to perform better, the Encoder-only model achieved notable results in a shorter timeframe, securing second place in the BraTS-METS 2023 challenge with a 59.8\% average lesion-wise Dice score on the test set.

\bibliographystyle{splncs04}
\bibliography{egbib}
\end{document}